\documentclass[aps,amsmath,onecolumn, showpacs,superscriptaddress,pre,10.5pt,nofootinbib]{revtex4-1}
\usepackage{amssymb}
\usepackage{amsmath}
\usepackage{graphicx}
\usepackage{array}
\usepackage{physics}
\usepackage{mathtools}
\usepackage{amssymb}
\usepackage{verbatim}
\usepackage{color}
\usepackage{bm}
\usepackage{bbm}
\usepackage{amsthm}
\usepackage[normalem]{ulem}
\usepackage{caption}
\usepackage[titletoc,title]{appendix}

\usepackage{color}
\usepackage[dvipsnames]{xcolor}
\definecolor{Blue}{rgb}{0.00, 0.00, 1.00}
\definecolor{Red}{rgb}{1.00, 0.00, 0.00}
\definecolor{Green}{rgb}{0.00, 0.50, 0.00}

\usepackage[colorlinks=true, urlcolor=blue, anchorcolor=blue, citecolor=blue,filecolor=blue,linkcolor=blue,menucolor=blue]{hyperref}

\newcommand{\bea}{\begin{eqnarray}}
\newcommand{\eea}{\end{eqnarray}}
\newcommand{\be}{\begin{equation}}
\newcommand{\ee}{\end{equation}}

\newcommand{\bee}{\begin{equation*}}
\newcommand{\eee}{\end{equation*}}

\colorlet{Mycolor1}{green!10!orange!90!}

\def\XXint#1#2#3{{\setbox0=\hbox{$#1{#2#3}{\int}$}
     \vcenter{\hbox{$#2#3$}}\kern-.5\wd0}}

\usepackage{listings}
\usepackage[T1]{fontenc}
\usepackage[latin9]{inputenc}
\usepackage{esint}
\usepackage{graphicx}
\usepackage{color}
\definecolor{Blue}{rgb}{0.00, 0.00, 1.00}
\definecolor{Red}{rgb}{1.00, 0.00, 0.00}
\definecolor{Green}{rgb}{0.00, 0.60, 0.00}

\usepackage{multirow}
\usepackage{float}
\usepackage{caption, subcaption}

\usepackage[colorlinks=true, urlcolor=blue, anchorcolor=blue, citecolor=blue,filecolor=blue,linkcolor=blue,menucolor=blue]{hyperref}

\begin{document}

\title{\bf Nonequilibrium steady state of Brownian motion in an intermittent potential}

\author{Soheli Mukherjee}
\email{soheli.mukherjee2@gmail.com}
\affiliation{Racah Institute of Physics, Hebrew University of Jerusalem, Jerusalem 91904, Israel}
\affiliation{Department of Environmental Physics, Blaustein Institutes for Desert Research, Ben-Gurion University of the Negev, Sede Boqer Campus, 8499000, Israel}

\author{Naftali R. Smith}
\email{naftalismith@gmail.com}
\affiliation{Racah Institute of Physics, Hebrew University of Jerusalem, Jerusalem 91904, Israel}
\affiliation{Department of Environmental Physics, Blaustein Institutes for Desert Research, Ben-Gurion University of the Negev, Sede Boqer Campus, 8499000, Israel}

\begin{abstract}

    We calculate the steady state distribution $P_{\text{SSD}}(\boldsymbol{X})$ of the position of a Brownian particle under an intermittent confining potential that switches on and off with a constant rate $\gamma$. We assume the external potential $U(\boldsymbol{x})$ to be smooth and have a unique global minimum at $\boldsymbol{x} = \boldsymbol{x}_0$, and in dimension $d>1$ we additionally assume that $U(\boldsymbol{x})$ is central.
    We focus on the rapid-switching limit $\gamma \to \infty$.
    Typical fluctuations follow a Boltzmann distribution $P_{\text{SSD}}(\boldsymbol{X}) \sim e^{- U_{\text{eff}}(\boldsymbol{X}) / D}$, with an effective potential   $U_{\text{eff}}(\boldsymbol{X}) = U(\boldsymbol{X})/2$, where $D$ is the diffusion coefficient. 
    However, we also calculate the tails of $P_{\text{SSD}}(\boldsymbol{X})$ which behave very differently. In the far tails $|\boldsymbol{X}| \to \infty$, a
    universal behavior 
    $P_{\text{SSD}}\left(\boldsymbol{X}\right)\sim e^{-\sqrt{\gamma/D} \, \left|\boldsymbol{X}-\boldsymbol{x}_{0}\right|}$
    emerges, that is independent of the trapping potential.
    The mean first-passage time to reach position $\boldsymbol{X}$ is given, in the leading order, by $\sim 1/P_{\text{SSD}}(\boldsymbol{X})$. This coincides with the Arrhenius law  (for the effective potential $U_{\text{eff}}$) for $\boldsymbol{X} \simeq \boldsymbol{x}_0$, but deviates from it elsewhere. 
    We give explicit results for the harmonic potential.
    Finally, we extend our results to periodic one-dimensional systems. Here we find that in the limit of $\gamma \to \infty$ and $D \to 0$,  the logarithm of $P_{\text{SSD}}(X)$ exhibits a singularity  
    which we interpret as a first-order dynamical phase transition (DPT). This DPT occurs in absence of any external drift. We also calculate the nonzero probability current in the steady state that is a result of the nonequilibrium nature of the system.

\end{abstract}

\maketitle

{%
	\hypersetup{linkcolor=black}
	\tableofcontents
}

\section{Introduction}





Stochastic resetting of random processes has gained considerable attention due to its remarkable properties. One of them is the emergence
of nonequilibrium behavior, and in particular, non-equilibrium steady states (NESS) that are reached at long times. In the ideal limit, stochastic resetting is instantaneous, and this assumption may be unrealistic for some systems.  For instance, the thermodynamic cost to change the particle's position instantaneously diverges. Due to this challenging nature of the setup the experimental  realization of the resetting is limited. The paradigmatic problem
of the one-dimensional Brownian motion with resetting to the origin \cite{EM11} has been experimentally realized with silica micro-spheres manipulated by optical tweezers \cite{BBPMC20}. In another experiment  the particle diffuses freely and after exponentially distributed
time intervals the particles are driven back to the starting position mimicking resetting events \cite{FPSRR20}.

To overcome this unsuitable feature of ideal resetting, some proposals have been introduced to mimic generic and physically realistic resetting systems, which in principle would not require to track a particle or return it to the origin in a controlled deterministic way: intermittent potentials \cite{MBMS20, SDN2021, MBM22, GPK2021, GPKP2021, GP2022, ACB2022, BKMS24} and  refractory periods \cite{EM2018}. The former  method considers an external trap, namely, a confining potential which is usually assumed to have a single minimum. 
The state of the potential is described by a time dependent binary variable $\eta(t)$, which switches on and off the potential $U(\boldsymbol{x})$ randomly.  Switching off the potential allows the free Brownian motion of the particle, while the on-state produces an attractive
motion of the particle towards the centre of the potential. These dynamics may be viewed as a non-ideal version of resetting of the particle's position to the minimum of the potential.

 The problem of Brownian particles under intermittent potentials has been studied in dimension $d=1$ for
confining potentials of the type $U(x) \propto |x|$ \cite{MBMS20, GPK2021, GPKP2021} and $U(x) \propto x^2$ (harmonic) \cite{SDN2021, Frydel24, GPKP2021, BKMS24}. The NESS (i.e  $P_{\text{SSD}}(X)$) and first passage properties were studied. The results were extended for a general confining potential of the form $U(x) = k |x-x_0|^n/n$ \cite{MBM22}. These works also investigated how the switching rates  from the on state to the off state and back (which, for simplicity, we will assume to be equal and denote them by $\gamma$)  affect the mean search time to reach a given target. In \cite{SDN2021}, the NESS was calculated exactly (in Fourier space) for  a harmonic intermittent potential in $d=1$. Moreover, its limiting behaviors were obtained for different parameter regimes (i.e $\gamma$ and the strength $\mu$ of the harmonic potential $U(x)$).
For arbitrary trapping potential $U(\boldsymbol{x})$, in the  limit $\gamma \to 0$, if one deletes from the dynamics the time intervals in which the potential is turned on, then the remaining dynamics are very similar to Brownian motion in presence of (instantaneous) resetting. This enables one to easily obtain the steady state distribution (SSD) using well-known results \cite{SDN2021}.
For intermediate values of $\gamma$, analytic progress is in general difficult (except in special cases \cite{SDN2021}).

Another experimentally realized limit is the limit of large switching rates $\gamma \to \infty$, which is very different from the resetting limit. 
This limit has been realized in the optical and acoustic trapping of passive and active particles \cite{TDVB2016, GAMHG2021, BCASL2022, GPMHG2022}. In this case the typical fluctuations of $P_{\text{SSD}}(\boldsymbol{X})$ follow a Boltzmann distribution near the centre of the trap, 
with an effective potential which is proportional to the original one \cite{SDN2021, Frydel24}. However, the effective thermal equilibrium approximation breaks down at sufficiently large values of $X =| \boldsymbol{X}|$. 
So the natural question that arises,  for $\gamma \to \infty$, is what happens when $\boldsymbol{X}$ is not close to the center of the trap, i.e how the non-equilibrium behaviour of the system affects the tail distribution of  $P_{\text{SSD}}(\boldsymbol{X})$.

 The latter question naturally leads us to the realm of large deviations (or rare events) \cite{DZ1998, Hollander2000, Hugo2009, Hugo2018},  a theme of ongoing interest in statistical mechanics. In general, large deviations may be of great importance in general despite their low likelihood, because they can have dramatic consequences (e.g., earthquakes, stock-market crashes etc). They are also of fundamental interest in statistical physics because they can display remarkable nonequilibrium effects.  Examples for such effects include significant departures from the Boltzmann distribution and Arrhenius law which typically describe systems in thermal equilibrium.

In this paper, we first consider a one-dimensional single Brownian motion subjected to an intermittent generic potential with a single global minimum at $x_0$. The potential switches on and off with a constant rate $\gamma$. 
We consider the limit of very fast switching $\gamma \to \infty$ (i.e, the timescale $1/\gamma$ is very small compared to the typical relaxation timescale associated with the potential $U(x)$).
We develop a general  large-deviations formalism for such systems, and apply it to calculate expressions for   the NESS and the MFPT to reach a given target for any confining potential. 
We uncover an interesting universal behavior in the far tails, $|X| \to \infty$. 
 We then extend our results to higher dimensions $d>1$ assuming rotational symmetry. As a demonstration, we perform explicit calculations for the harmonic potential (in general dimension $d$).
Finally, we study one-dimensional periodic systems and find that, at $\gamma \to \infty$ and $D \to 0$, the  large-deviation function (LDF) that describes the NESS is in general singular. In analogy with statistical mechanics in equilibrium, we interpret this singularity as a dynamical phase transition (DPT) \cite{Baek15, Baek17, Baek18,  Shpielberg2016, ZM16, SMS2018, ALM19, Agranov2020, Smith22Chaos, ACJ22, SGG23, ACJ23, SSM23, MLS24, Shpielberg24} of the first order.

 Our theoretical formalism is based on two fundamental steps, and each one of these steps is based on a different standard theoretical tool from large-deviation theory. First, we coarse grain the dynamics over timescales much larger than $1/\gamma$ by using the long-time large-deviation principle. This can be done, e.g.,  by using Donsker-Varadhan theory \cite{Hugo2018}. Next, we apply the Optimal Fluctuation method (OFM) \cite{Onsager, MSR, Freidlin, Dykman, Graham, Falkovich01, GF, EK04, Ikeda2015, Grafke15, MeersonSmith19, SmithMeerson19, Meerson2023, NF2022, NS2023, VLM24} on the coarse-grained effective dynamics.
The OFM is also known in other contexts by different names, such as weak-noise theory or the instanton method. It yields the (approximate) probability of the large deviation by finding the optimal (i.e., most likely) history of the system conditioned on the occurrence of the rare event -- and this is an interesting physical observable in its own right.
Theoretical approaches similar to the one we use here (i.e., applying the OFM to coarse-grained dynamics) have been applied successfully in a variety of physical systems.
In some contexts the approach is known as the ``temporal additivity principle'' \cite{NF2022, NS2023, DZ1998, Hollander2000, Hugo2009, HT2009, Harris2015, Jack2019, JH2020, AB2021, NF2022, NS2023, BTZ2023, MS2024, PLC24, DKB25}.

The rest of the paper is organised as follows. 
 In Sec. \ref{sec model} we introduce the model of a Brownian particle in an intermittent potential in $d=1$.  In Sec. \ref{sec formalism}, we develop the formalism to calculate the NESS for any general potential in $d=1$ in the limit $\gamma \to \infty$, following the two steps described above. We then extend the results to $d>1$ assuming rotational symmetry.
In Sec.~\ref{sec harmonic}, we perform the calculations explicitly for a harmonic confining potential in general dimension $d$. In Sec. \ref{sec generalization} we consider one-dimensional periodic potentials, uncover a first-order DPT  in the large-deviation function that describes the SSD, and calculate the nonequilibrium current in the steady state. Finally, we summarize and discuss our results in Sec. \ref{sec conclusion}. 

\bigskip

\section{Model} \label{sec model}

We consider a Brownian motion in a one-dimensional ($d=1$)  potential $U(x)$ which is intermittent in time. The Langevin equation describing the dynamics is given by the following equation:
\begin{equation}
\dot{x}=F\left(x\right)\eta\left(t\right)+\sqrt{2D}\,\xi\left(t\right)
\end{equation}
where $\xi\left(t\right)$ is a Gaussian white noise with zero mean and delta-correlated two-point
correlation $\langle \xi(t) \, \xi(t') \rangle =  \delta(t-t')$, $D$ is the diffusion constant and $\eta(t)$ switches between the off and on states (with values 0 and 1 respectively) stochastically with the rate $\gamma$ (where $\tau = \gamma^{-1}$ is the  mean switching time). 
$F(x)$ is the external force extracted from the potential $U(x)$:
 \begin{equation}
F\left(x\right) = - U'\left(x\right).
 \end{equation}
Here for simplicity, we assume $U(x)$ has a single global minimum at $x=x_0$, with no additional minima or maxima. At long times, the position of the particle will approach a SSD \cite{SDN2021}. There are two important timescales in this system. One is $1/\gamma$ and another one is the typical time taken by the particle to relax to  the point $x=x_0$  when the potential is on. We aim to calculate the full  SSD $P_{\text{SSD}}(X)$ of the particle 
in the limit where $\eta\left(t\right)$ switches very quickly ($\gamma \to \infty$ or $\tau \to 0$). In this limit, the rapid switching of the noise causes it to typically average out to value $\left\langle \eta\left(t\right)\right\rangle =1/2$. Hence, for typical fluctuations (when $X$ is close to $x_0$) we can approximate
\begin{equation} \label{Effective Langevin}
    \dot{x}\simeq F\left(x\right)/2+\sqrt{2D}\,\xi\left(t\right).
    \end{equation}
As a result, the system effectively reaches thermal equilibrium, so the SSD is given by \cite{SDN2021}
    \begin{equation} \label{typical fluc}
    P_{\text{SSD}}\left(X\right)\simeq Z^{-1} e^{-U_{\text{eff}}\left(X\right)/D},\quad \text{where} \quad U_{\text{eff}}\left(X\right)= \frac{1}{2} \, \Big(U\left(X\right) - U(x_0)\Big )
\end{equation}
and $Z$ ensures the normalization.

However, atypical values of $X$ are associated with unusual realizations of $\eta(t)$,  with time averages which differ significantly from the average value $1/2$.
Our goal is  to calculate the tails of the distribution (large $|X|$).
It is convenient to rewrite the Langevin equation in terms of  telegraphic noise $\sigma(t) = 2\eta(t)-1$
\be \label{SDE}
    \dot{x}=\frac{F\left(x\right)}{2}+\frac{F\left(x\right)}{2}\sigma\left(t\right)+\sqrt{2D}\,\xi\left(t\right) \, ,
\ee
where $\sigma\left(t\right)=\pm1$ (and switches very rapidly, at rate $\gamma$) has zero mean. Eq. \eqref{SDE} looks similar to the Langevin equations for trapped run-and-tumble (active) particles \cite{TC2008, TC2009} but in our case the telegraphic noise is multiplicative instead of additive, and there is  an additional white noise term.
In the limit of rapid switching, the tails of the SSD for trapped run-and-tumble and other active particles have been amenable to theoretical analysis using a coarse-grained OFM approach \cite{NF2022, NS2023}.
The SSD was found to be related to the 
particle's position distribution at long-times in the absence of an external potential. 
In the next section, we will develop a similar formalism to calculate the SSD for Brownian particles trapped in intermittent potentials.

\bigskip

\section{Formulation: Coarse-graining and optimal fluctuations method} \label{sec formalism}

Our theoretical framework first consists of a coarse graining of the dynamics Eq. \eqref{SDE} over intermediate timescales, that are much larger than $1/\gamma$ but still shorter than the typical relaxation timescale of the system.  
Then we calculate the most likely coarse-grained trajectory $x(t)$ by applying the OFM to the path-integral formulation of the coarse-grained dynamics similar to \cite{HT2009, Harris2015, Jack2019, JH2020, AB2021, NF2022, NS2023, BTZ2023, MS2024, PLC24, DKB25}. $P_{\text{SSD}}(X)$ is then approximated by the probability of the optimal path that leads the particle from an initial relaxation state to the point $X$.

\subsection{Coarse-graining the dynamics} \label{sec coarse grain}


In order to perform the coarse graining, we first treat the simplest case when $F(x) = f$ is a constant.
In this case the system never reaches a steady state, but its dynamical fluctuations can be analyzed as follows. 
For this case of a constant force, let us denote the particle's position by $\tilde{x}(t)$, and assume that the particle starts at the origin $\tilde{x}(0)=0$. Plugging $F(x)=f$ into the dynamics in Eq.~\eqref{SDE}, we can formally solve for the time-dependent position of the particle
\begin{eqnarray} \label{coarsegrainedSDE const f}
    \tilde{x}(t)=\frac{f \, t}{2}+\frac{f}{2} \int_0^t \, \sigma \left(t'\right) \, dt' + \sqrt{2D}\, \int_0^t \, \xi\left(t'\right) \, dt'.
\end{eqnarray}
In the long time limit, the distribution of $\tilde{x}$ follows a large deviation principle (LDP) \cite{Hugo2018, BMRS2019, SBS2020, DMS2021, NF2022, NS2023} 
\be \label{proboftrajectory}
P[\tilde{x}(t)] \sim e^{- t\,  \Psi_f(\tilde{x}/t)} \, .
\ee
Here $ \Psi_f(v)$ is the rate function and $v=\tilde{x}/t$ is the empirical velocity of the particle in the time interval $[0,t]$.  $ \Psi_f(v)$ can be calculated from the Legendre-Fenchel transformation of the  scaled cumulant generating function (SCGF) 
$\lambda_{f}\left(k\right)= \displaystyle {\lim_{t\to\infty} \frac{1}{t} \, \ln\langle e^{k\tilde{x}}\rangle}$
of the position of the particle \cite{Hugo2018}, which depends on $f$. 
It is easier to work with the SCGF $\lambda_f$, since the cumulants of the sum of independent random variables are the sum of their cumulants.
As a result,  $\lambda_f$ is given by the sum of the SCGFs of all the terms in the RHS of Eq. \eqref{coarsegrainedSDE const f}.
These SCGFs are all easy to calculate and well known (see Appendix \ref{SCGFcalculation} for details) \cite{SBS2020, NF2022, NS2023}, and we obtain
 \be \label{SCGF}
     \lambda_f(k) = \frac{f \, k}{2} + \sqrt{\frac{f^{2}k^{2}}{4}+\gamma^2} -\gamma +D\, k^{2} \, .
 \ee
$\Psi_f(v)$ can be now calculated by the Legendre-Fenchel transformation of 
$\lambda_{f}(k)$: $\Psi_{f}(v)=\sup_{k}{\displaystyle \lbrace v\,k-\lambda_{f}(k)\rbrace} $.
Since $\lambda_f$ is differentiable and convex, the Legendre-Fenchel transform reduces to the Legendre transform 
\cite{TouchetteNutshell}. This enables us to calculate $\Psi_{f}(v)$ in a parametric form
  \bea
  \label{LF calculation}
    v(k) &=&\frac{d \lambda_f}{d k}=\frac{f}{2}+2\, D\, k+\frac{f^{2}k}{2\sqrt{4 \gamma^2+f^{2}k^{2}}} \, , \\
 \label{Psi general}
\Psi_{f}(k)  &=&  k \, v - \lambda_f= \gamma + D\, k^{2}-\frac{2 \, \gamma^2}{ \sqrt{4 \gamma^2 + f^{2}k^{2}}} \, .
\eea

We next use this result for $F(x) = f$ as a starting point to continue the calculations for  $F(x) \neq $ constant (Eq. \eqref{SDE}). In this case one can coarse grain the dynamics (this is essentially equivalent to coarse graining the telegraphic noise $\sigma(t)$)  over time windows much longer than $1/\gamma$. 
We can nevertheless choose these time windows to be much shorter than the system's relaxation timescale, thus exploiting the timescale separation.
One can then approximate $F(x)$ to be constant within each time window, and also approximate the noise terms to be statistically independent between different time windows.
As a result, we get using Eq. \eqref{proboftrajectory} the probability of a coarse-grained trajectory $x(t)$ of duration $T$ as
\begin{eqnarray}
\label{PxtCoarseGrained}
P\left[x\left(t\right)\right] 
    \sim e^{-\int_{0}^{T_{1}}\Psi_{f_1}\left(\dot{x}\right)dt}e^{-\int_{T_{1}}^{T_{2}}\Psi_{f_2}\left(\dot{x}\right)dt}e^{-\int_{T_{2}}^{T_{3}}\Psi_{f_3}\left(\dot{x}\right)dt}...\sim e^{-\int_{0}^{T}\Psi_{F\left(x\right)}\left(\dot{x}\right)\, dt},
    \end{eqnarray} 
    where $0 < T_1 <T_2 < \dots$ define the time windows and their durations are $T_{i+1}-T_i$, and $f_i = F(x(T_i))$ is the approximately constant value of the external force in the $i$th interval. 
    A similar coarse-graining procedure has been used in many physical systems in which there is timescale separation, and in some contexts it is referred to as the temporal additivity principle \cite{NF2022, NS2023, DZ1998, Hollander2000, Hugo2009, HT2009, Harris2015, Jack2019, JH2020, AB2021, NF2022, NS2023, BTZ2023, MS2024, PLC24, DKB25}.
    
    The procedure that we have performed is essentially equivalent to coarse-graining the telegraphic noise term, leading to the coarse-grained dynamics
\be \label{coarsegrainedSDE}
    \dot{x}(t)=\frac{F\left(x\right)}{2}+\frac{F\left(x\right)}{2}\bar{\sigma}\left(t\right)+\sqrt{2D}\,\xi\left(t\right) \, ,
\ee
where
\be \label{observable}
\bar{\sigma}(t) = \frac{1}{\mathfrak{T}} \, \int_0^\mathfrak{T} \, \sigma(t+t') \, dt'
\ee
is the time average of the telegraphic noise over a window of intermediate duration  $\mathfrak{T}$. Using a similar procedure to the one above, we can write the probability for a coarse-grained trajectory $\bar{\sigma}(t)$  of duration $T$ as
\begin{eqnarray} \label{pdf of sigma}
P\left[\bar{\sigma}\left(t\right)\right]\sim e^{- \gamma \int_{0}^{T}\psi\left(\bar{\sigma}\left(t\right)\right)dt},
\end{eqnarray}
where $\psi\left(z\right)= 1-\sqrt{1-z^{2}}$ with $-1 \leq z \leq 1$ has been known, e.g., from the context of the position distribution of a run-and-tumble particle at long times \cite{SBS2020, NF2022, NS2023}. 
    
\smallskip

We now use Eq. \eqref{PxtCoarseGrained} as the starting point for the application of  OFM to calculate the optimal (coarse-grained) history of the system conditioned on a given rare event. The OFM is based on a saddle-point evaluation of the path integral of the (coarse-grained) process. 
Let us rewrite Eq. \eqref{PxtCoarseGrained} in the form
\be 
P[x(t)] \sim e^{- S[x(t)]}
\ee
where the action functional $S[x(t)]$ is given by
\begin{equation} \label{action}
   S[x(t)] = \int_0^T \Psi_{F(x)}(\dot{x}) \,  dt
\end{equation}
and  $\Psi_{F\left(x\right)} \! \left(\dot{x}\right)$ is given in a parametric form by Eqs.~\eqref{LF calculation} and \eqref{Psi general} with $f$ replaced by $F(x)$. For future convenience, let us give these equations explicitly
\bea
\label{lambdaFprime}
\dot{x} &=& \lambda'_F(k) = \frac{F(x)}{2}+2\, D\, k+\frac{F(x)^{2}k}{2\sqrt{4 \, \gamma^2 +F(x)^{2}k^{2}}} \, , \\
\Psi_{F(x)}(k) &=&  \gamma + D\, k^{2}-\frac{2 \, \gamma^2}{ \sqrt{4 \, \gamma^2 +F(x)^{2}k^{2}}} \, .
\label{ratef F}
\eea 
We also give here, for future convenience, the explicit formula for $\lambda_{F(x)}(k)$:
\be \label{CGF general F}
 \lambda_{F(x)}(k) = \frac{F(x) \, k}{2} + \sqrt{\frac{F(x)^{2}k^{2}}{4}+\gamma^2} -\gamma +D\, k^{2} \, .
\ee

\subsection{Finding the optimal coarse-grained path} \label{sec optimal path}

We are now interested in calculating $P_{\text{SSD}}(X)$. We therefore assume that the system has had a very long time to evolve prior to the measurement time. 
In principle, one can calculate $P_{\text{SSD}}(X)$ by evaluating the path integral that corresponds to the action \eqref{action} over the trajectories $x(t)$ of very long duration, that begin from a relaxation state and end at the point $X$.
It is convenient to take the time window for these trajectories to be $(-\infty,0]$, and then the constraints become $x(-\infty) = x_0$ and $x(0)=X$.
In the large-deviations regime, corresponding to unlikely values of $X$,  the action $S[x(t)]$ becomes large and we can evaluate the path integral using the saddle-point approximation.
This now leads to a minimization problem for the action Eq. \eqref{action}
    over the trajectories $x(t)$ which is to be solved subject to the  boundary conditions
\be
x(t\to -\infty) = x_0, \qquad x(t=0) = X . 
\ee


We will now develop an analogy to a classical Hamiltonian system. We observe that  Eq.~\eqref{action} has the form of a least action principle
where  
$L(x, \dot{x})=\Psi_{F\left(x\right)}\left(\dot{x}\right)$
is the Lagrangian of underlying effective Hamiltonian dynamics. 
 Since the Legendre transform of 
$L(x, \dot{x})=\Psi_{F\left(x\right)}\left(\dot{x}\right)$
is given by $\lambda_{F(x)}(k)$, we identify the latter as the effective Hamiltonian $H(x, k)$ of the system. Here $k$ plays the role of conjugate momentum to $x$, and is related to $x$ and $\dot{x}$ through  \eqref{lambdaFprime}. Furthermore, since
 $L$ does not depend on $t$ explicitly, the corresponding  Hamiltonian is conserved
\be \label{Hamiltonian}
H(x,k)=\dot{x} \,\,\frac{\partial L}{\partial\dot{x}}-L=\lambda_{F(x)}(k)= \text{const}=E.
\ee
The constant in \eqref{Hamiltonian} can be found from the boundary condition at $t\to-\infty$. 
At time $t\to-\infty$, $\dot{x}=0$ and $x=x_{0}$ (the minimum of the potential) so also $F=0$. Using Eq.   \eqref{lambdaFprime}, the $k$ becomes $k(t=-\infty)=0$. Hence the constant is $E = 0$. 
 Using this we find that the action, evaluated on the optimal path, is given by
\begin{eqnarray}
\label{action calculation}
S\left(X\right)	= \int_{-\infty}^{0} (\dot{x} \, k - H(x,k))dt
 =  \int_{x_{0}}^{X}k \,dx
 \end{eqnarray}
 where $k$ and $x$ are related through $\lambda_{F(x)}(k) = 0$ and \eqref{CGF general F}, which together simplify to:
\begin{eqnarray} \label{relation of k and x}
\frac{2\, D\, k\, \gamma-D^2\,k^{3}}{D\, k^{2}-\gamma}&=F\left(x\right).
\end{eqnarray}
  The solution to Eq.~\eqref{relation of k and x}  can be written in terms of the dimensionless parameter $w = \frac{F(x) }{\sqrt{D \, \gamma}}$ as:
\be \label{kF(x) relation}
\sqrt{D /\gamma}\,k=\Omega\left(\frac{F\left(x\right)}{\sqrt{D \, \gamma}}\right)
\ee
where 
\be \label{mu}
\Omega(w)=\frac{1}{3}\left(-\frac{w^{2}+6}{\beta}-\beta-w\right);\quad\beta=e^{i\pi/3}\sqrt[3]{-w^{3}+\frac{3\sqrt{3}}{2}\sqrt{-4w^{4}-13w^{2}-32}+\frac{9}{2}w}
\ee
 is the solution to the cubic equation 
 \be
 \label{OmegaCubicEqn}
  \Omega^3+ w\,  \Omega^2 -2 \, \Omega - w=0,
  \ee
 and $\sqrt[3]{\dots}$ denotes the principal complex cube root.
 Finally, the SSD is given by
 \be \label{SSD small tau}
 P_{\text{SSD}}(X)\sim e^{-S(X)}
 \ee
and the  the MFPT is $\mathcal{
T} \sim 1 / P_{\text{SSD}} (X) \sim  e^{S(X)}$,
where $S(X)$ is given by Eqs.~\eqref{action calculation}, \eqref{kF(x) relation} and \eqref{mu}. These formulas for the SSD and MFPT constitute a central result of this paper, and they are valid for general trapping potentials (in $d=1$ and under the mild assumptions mentioned above,  namely, that the potential is smooth and has a single minimum, which is at $x=x_0$).

The asymptotic behaviour of the action for small and large $X$ can be extracted from the asymptotic behaviours of $\Omega(w)$:
\be \label{Omega asymp}
\Omega(w)=\begin{cases}
-\frac{w}{2}+O\left(w^{3}\right), & |w|\ll1,\\[1mm]
-\text{sgn}(w)+\frac{1}{2\,w}+O\left(\frac{1}{w^{2}}\right), & |w|\gg1.
\end{cases}
\ee
Therefore the asymptotics of the action are given by 
\be
S(X) \simeq \begin{cases}
- \frac{1}{2 D}  \int_{x_0}^X  F(x) \, dx  = \frac{U(X)}{2D}, & X \simeq x_0,\\[2mm]
\sqrt{\frac{\gamma}{D}} \, |X-x_0|,
& |X| \to \infty,
\end{cases}
\ee
 where, in the second line of the equation, we assumed (in addition to the general assumptions made above) that as $X\to  \pm \infty$, so does $F(X)\to  \mp \infty$, i.e., that the potential is sufficiently strongly confining.
The asymptotic behavior of $S(X)$ for $X \simeq x_0$ matches smoothly with the regime of typical fluctuations given by Eq. \eqref{typical fluc}.
Note also that the leading-order behavior at large $|X|$,  corresponding to an exponential tail
$P_{\text{SSD}}(X)\sim e^{-\sqrt{\gamma/D}\,\left|X-x_{0}\right|}$,
is universal and independent of $U(X)$,
 and therefore, it in general differs by many orders of magnitude from the tail of the effective Boltzmann distribution \eqref{typical fluc}. The physical mechanism that leads to this universality will become clear shortly.
 This exponential tail coincides, in the leading order, with the nonequilibrium steady-state distribution of the position of a Brownian particle with instantaneous resetting to $x_0$ at a constant rate $\gamma$ \cite{EM11}. We will uncover the reason for this coincidence below.
Incidentally, it is interesting to note that universal exponential tails have been found in other classes of systems too, e.g., in models of continuous-time random walks or in models of diffusing diffusivity \cite{YAMM2021, CSMS2017, BB20, WBB20, HuEtAl23, Burov20, SB23, HBB24, SB24} as well as in experiments \cite{CK20, SIJ23}.

\smallskip

Our coarse grained method enables us to calculate, in addition to the SSD itself, also the optimal (most likely) coarse-grained trajectory of the system conditioned on a given value of $X$. 
The optimal path that minimizes $S(X)$ can be calculated using Eq. \eqref{kF(x) relation} and Eq. \eqref{mu}. 
 Plugging Eq.~\eqref{kF(x) relation} into \eqref{lambdaFprime}, we obtain
\be
\dot{x}= \lambda_{F\left(x\right)}'\left(  g(x)\right)
\ee
where $g(x)=\sqrt{\frac{\gamma}{D}}\,\,\Omega\Big(\frac{F\left(x\right)}{\sqrt{D \, \gamma}} \Big) $,  and so we obtain the optimal path $x(t)$ in the form:
\be \label{op trajec}
t\left(x\right)=\int_{X}^{x}\frac{dx'}{\dot{x}'}=\int_{X}^{x}\frac{dx'}{\lambda_{F\left(x'\right)}'\left(g(x')\right)} \, .
\ee

If we assume again that at   $x \to \pm \infty$, $F(x) \to \mp \infty$,  we find a universal optimal path, as we now show. In that case, for sufficiently large $|X|$, the strong inequality $4\,\gamma D\ll F(x)^{2}$ holds, and we can use the asymptotic behavior \eqref{Omega asymp} of $\Omega$ to obtain $k \simeq \text{sgn}(X)\sqrt{\gamma / D }$. Using this in \eqref{lambdaFprime} (and again using $4\,\gamma D\ll F(x)^{2}$) we obtain
\be
\lambda'_F(k)  \simeq \text{sgn}(X) \sqrt{4 D \, \gamma} \, .
\ee
Using \eqref{op trajec}, we then find that the optimal path can be approximated as:
\be
t\left(x\right)\simeq-\left|\int_{X}^{x}\sqrt{\frac{1}{4\,D\,\gamma}}\,d\tilde{x}\right|=-\sqrt{\frac{1}{4\,D\,\gamma}}\,\left|X-x\right| \, .
\ee 
This is valid only for $x$ that is not close to $x_0$, so in fact we get
\be \label{optimal path large X}
x\left(t\right)\simeq\begin{cases}
x_{0}, & t<T,\\[2mm]
\frac{X\,T+(x_{0}-X)\,t}{T}\,, & T<t<0,
\end{cases}
\ee
with $T=-\left|X-x_{0}\right|/\sqrt{4D\gamma}$.

To find the optimal path in the opposite limit $X \simeq x_0$, we 
 expand \eqref{CGF general F} at small $F(x)$
\be
H\left(x,k\right)=\lambda_{F\left(x\right)}\left(k\right)\simeq k\left[\frac{F\left(x\right)}{2}+Dk\right]
\ee
which corresponds to the Lagrangian
\be
L\left(x,\dot{x}\right)\simeq\frac{1}{4D}\left(\dot{x}-\frac{F\left(x\right)}{2}\right)^{2} \, .
\ee
This Lagrangian is that of the  OFM action for the Langevin equation given in Eq. \eqref{Effective Langevin}
which describes the motion of a Brownian particle in the effective (non-intermittent) potential $U(x)/2$, 
 which again would yield that typical fluctuations are given by the effective Boltzmann distribution \eqref{typical fluc}. Moreover, the optimal path $x(t)$ in this case is the  time-reversed relaxation (noise-free) trajectory, $x(t) = x_{\text{rel}}(-t)$ where $x_{\text{rel}}(t)$ is defined for $t \ge 0$ and it evolves according to the deterministic (noise-free) equation:
\be \label{activation trajec}
\dot{x}_{\text{rel}}= F\left(x_{\text{rel}}\right)/2
\ee
 with the initial condition $x_{\text{rel}}(0) = X$, eventually relaxing to $x_{\text{rel}}(\infty) = x_0$.

\smallskip
In addition to the optimal trajectory $x(t)$ conditioned on $X$, our formalism also enables us to calculate the corresponding optimal realizations of the noise terms $\xi$ and $\bar{\sigma}$.
For that, we use the distributions \eqref{pdf of sigma} and
$P\left[\xi\left(t\right)\right]\sim e^{-\frac{1}{2}\,\int_{0}^{T}\xi^{2}\left(t\right)dt}$ 
of the coarse-grained and white noises respectively.
We can thus find the optimal realization of $\bar{\sigma}(t)$ and $\xi(t)$ conditioned on a given trajectory $x(t)$.
Isolating $\xi$ from the equation of motion Eq. \eqref{coarsegrainedSDE}, yields
\be \label{optimal xi}
\xi\left(t\right)=\frac{\dot{x}-\frac{F\left(x\right)}{2}-\frac{F\left(x\right)}{2}\bar{\sigma}\left(t\right)}{\sqrt{2D}},
\ee
and thus the conditional distribution of a realization $\bar{\sigma}(t)$ on  a given trajectory $x(t)$ is
\be
P\left[\bar{\sigma}\left(t\right)\,|\,x\left(t\right)\right]\sim e^{-\gamma \,\int_{0}^{T}\Psi\left(\bar{\sigma}\left(t\right)\right)dt}e^{-\frac{1}{2}\,\int_{0}^{T}\xi^{2}dt}=e^{-\int_{0}^{T}\left\{ \gamma \, \Psi\left(\bar{\sigma}\left(t\right)\right)+\frac{1}{4D}\left[\dot{x}-\frac{F\left(x\right)}{2}-\frac{F\left(x\right)}{2}\bar{\sigma}\left(t\right)\right]^{2}\right\}dt } \, .
\ee
We need to minimize the integral with respect to $\bar{\sigma}(t)$, but this is a simple problem: We simply take the derivative of the integrand with respect to $\bar{\sigma}(t)$ to obtain
\be \label{optimal noise}
\gamma \, \Psi'\left(\bar{\sigma}\left(t\right)\right)+\frac{1}{2D}\frac{F\left(x\right)}{2}\left[\frac{F\left(x\right)}{2}\bar{\sigma}\left(t\right)-\dot{x}+\frac{F\left(x\right)}{2}\right]=0 \, .
\ee
 The optimal realization of the coarse-grained telegraphic noise is given by solving this equation for $\bar{\sigma}(t)$ (given the optimal $x(t)$ and $\dot{x}(t)$ which are already known), and then using Eq.~\eqref{optimal xi} we get the optimal realization of the white noise $\xi(t)$ too.

 For typical fluctuations $X \simeq x_0$, the effective Langevin dynamics are given by Eq.~\eqref{Effective Langevin} and the optimal path is given by the time-reversed relaxation trajectory. Therefore, the optimal realization of the noises is
\be \label{opt noise large X}
\bar{\sigma} \simeq 0 \quad \text{and} \quad
\xi(t) \simeq -F\left(x\left(t\right)\right)/\left(2\sqrt{2D}\right).
\ee
The telegraphic noise $\bar{\sigma}$ averages out zero ($\eta(t)$ averages to 1/2) and the dominant contribution to the fluctuation is due to the white noise $\xi(t)$.

For $|X| \to \infty$ (assuming $F(x\to \pm \infty)\to \mp \infty$), using Eq. \eqref{optimal path large X}  with \eqref{optimal noise}, the optimal realization of the noises are 
\be \label{opt noise small X}
\begin{cases}
\bar{\sigma}(t) \simeq 0,\quad \;\; \xi(t)= 0 \, , \, & t<T,\\[1mm]
\bar{\sigma}(t) \simeq - 1\,, \quad \xi(t)= \text{constant}= \text{sgn}(X) \sqrt{2 \, \gamma} \, , & T<t<0.
\end{cases}
\ee
where $T=-|X - x_0| /\sqrt{4 \, D \, \gamma}$.  In other words, for $t<T$, the noises average out to zero and hence the particle stays $x \simeq x_0$. For $T < t < 0$, 
the trapping potential is in the ``off'' state $\eta = 0$ (corresponding to $\bar{\sigma}=-1$) and the white noise drives the particle from $x_0$ to $|X|$ at a constant velocity, as described by Eq. \eqref{optimal path large X}.
These results are universal, i.e., independent of the trapping potential under the assumptions given above.

 As mentioned above, the universal exponential $|X| \to \infty$ tail of the SSD coincides, in the leading order, with the SSD for a Brownian motion with instantaneous resetting to position $x=x_0$ at rate $\gamma$ \cite{EM11}. The physical mechanism behind this coincidence becomes transparent within the theoretical framework of the OFM. For the resetting Brownian motion, the optimal path leading to position $X$ at time $t=0$ is also given by Eq.~\eqref{optimal path large X} and with the same value of $T$. For this optimal path, no resetting events occur in the time interval $T<t<0$, while the corresponding optimal realization of the thermal (white) noise is given by the same constant from the second line in Eq.~\eqref{opt noise small X}. The probabilities for the optimal paths in the two problems coincide in the leading order, and from here it immediately follows that the (exponential) tails of the two SSD's coincide as well.


\subsection{Extension to higher dimensions}

The above formalism can be extended to higher dimensions $d>1$. The action given by Eq. \eqref{action} in the higher dimension becomes
\begin{equation} 
   S[\boldsymbol{x}(t)]=\int_{0}^{T}\Psi_{\boldsymbol{F}\left(\boldsymbol{x}\right)}(\dot{\boldsymbol{x}})\,dt
\end{equation}
 where the Legendre transform (in $d$ dimensions) of $\Psi_{\boldsymbol{F}\left(\boldsymbol{x}\right)}\left(\dot{\boldsymbol{x}}\right)$ is calculated similarly to the case $d=1$.
In the limit $\gamma \to \infty$, one can obtain the NESS by minimizing this action under appropriate temporal boundary conditions (as we did above for the case $d=1$).
In general, this minimization problem may be very difficult, however
it becomes very straightforward to find the solution of the problem when the system has rotational symmetry. For rotationally symmetric potential $U(\boldsymbol{r})=U(r)$ and assuming that the minimum of $U(r)$ is at the origin, 
the corresponding SSD $P_{\text{SSD}}(\boldsymbol{X})$ also becomes rotationally symmetric $P_{\text{SSD}}(\boldsymbol{X})=P_{\text{SSD}}^{\left(1D\right)}\left(X\right)$.
Moreover, the optimal path $\boldsymbol{x}(t)$ that minimizes $S[\boldsymbol{x}(t)]$ in this case follows a straight line joining between the origin and the point $\boldsymbol{X}$.
This reduces the problem to an effective one-dimensional one, for which the rate function $\Psi_{\boldsymbol{F}}(\boldsymbol{z})$ is replaced by  $\Psi_F(z)$ (given by Eqs.~\eqref{lambdaFprime} and \eqref{ratef F}). 
And the SSD $P_{\text{SSD}}^{\left(1D\right)}\left(X\right)$ is approximately that of the effective model in $d=1$ with potential $U(|x|)$. 
A very similar situation occurs in other nonequilibrium systems, e.g. for active particles confined by trapping potentials in the limit where the microscopic correlation time is small \cite{MDKD2020, Frydel2022, SDMS2022, NS2023}.

Let us now illustrate the general results obtained in the present section by considering the particular example of the harmonic trapping potential.

\section{Harmonic potential} \label{sec harmonic}

In this section, we show the explicit results for a Brownian particle confined in a $d$ dimensional harmonic potential $U(\boldsymbol{x}) = \frac{1}{2} \, \mu \, x^2$. For $d>1$, following the argument from the end of the previous section, due to the rotational symmetry, the SSD coincides in the leading order with that of the corresponding problem in $d=1$ with the potential $U(x) = \frac{1}{2} \, \mu \, x^2$. Therefore, in the remainder of the section we will treat the $d=1$ case.
As elsewhere in the paper, we consider the rapid switching limit. For the harmonic potential, this limit is given by $\gamma \gg \mu$, since this is the condition that ensures that the timescale $\mu^{-1}$ for relaxation to the minimum of the potential in the absence of noise  is  much larger than the switching time $\gamma^{-1}$. 
  Note that $U(x)$ satisfies the general assumptions from the previous section: It is a smooth function of $x$ and has a single global minimum at $x_0=0$. The corresponding force is $F(x) = - U'(x) = - \mu \, x$. For $x \to \pm \infty$, $F(x) \to \mp \infty$, i.e, it is strongly confining. 
  The exact SSD (that is reached at long times) was calculated in Fourier space in \cite{SDN2021}. 

In order   to calculate the SSD,
we start with the expression given by Eq. \eqref{CGF general F} which, for $F(x) = -\mu x$, becomes
\begin{eqnarray}
\lambda_{F(x)}\left(k\right)= - \frac{\mu \, x k}{2} + \sqrt{\frac{\mu^2 x^{2}k^{2}}{4}+\gamma^2} - \gamma  + D\, k^{2} \, .
\end{eqnarray}
 In the remainder of this section, we use units of length and time such that $D = \mu = 1$. Moreover, it is convenient to use the following scaled variables: $k = \sqrt{\gamma} \, \tilde{k}$
and $x = \sqrt{\gamma} \, y$, such that 
the action Eq. \eqref{action calculation} becomes 
\begin{eqnarray}
\label{SYHarmonic1}
    S(X, \gamma, 1)= \gamma \,  \mathcal{S}(Y) = \gamma \int_{0}^{Y} \, \tilde{k}(y) \,\, dy
\end{eqnarray}
where $Y=X / \sqrt{\gamma}$  and $\tilde{k}(y) = \Omega(-y)$ in Eq. \eqref{mu}. $\mathcal{S}(Y)$ is the large-deviation function that describes the SSD at $\gamma \to \infty$ through the LDP
\begin{eqnarray}
    P_{\text{SSD}}\left(X\right)\sim e^{-\gamma \, \mathcal{S}\left(X/\sqrt{\gamma}\right)} \, .
\end{eqnarray}

\begin{figure}
    \centering 
     \includegraphics[scale=0.7]{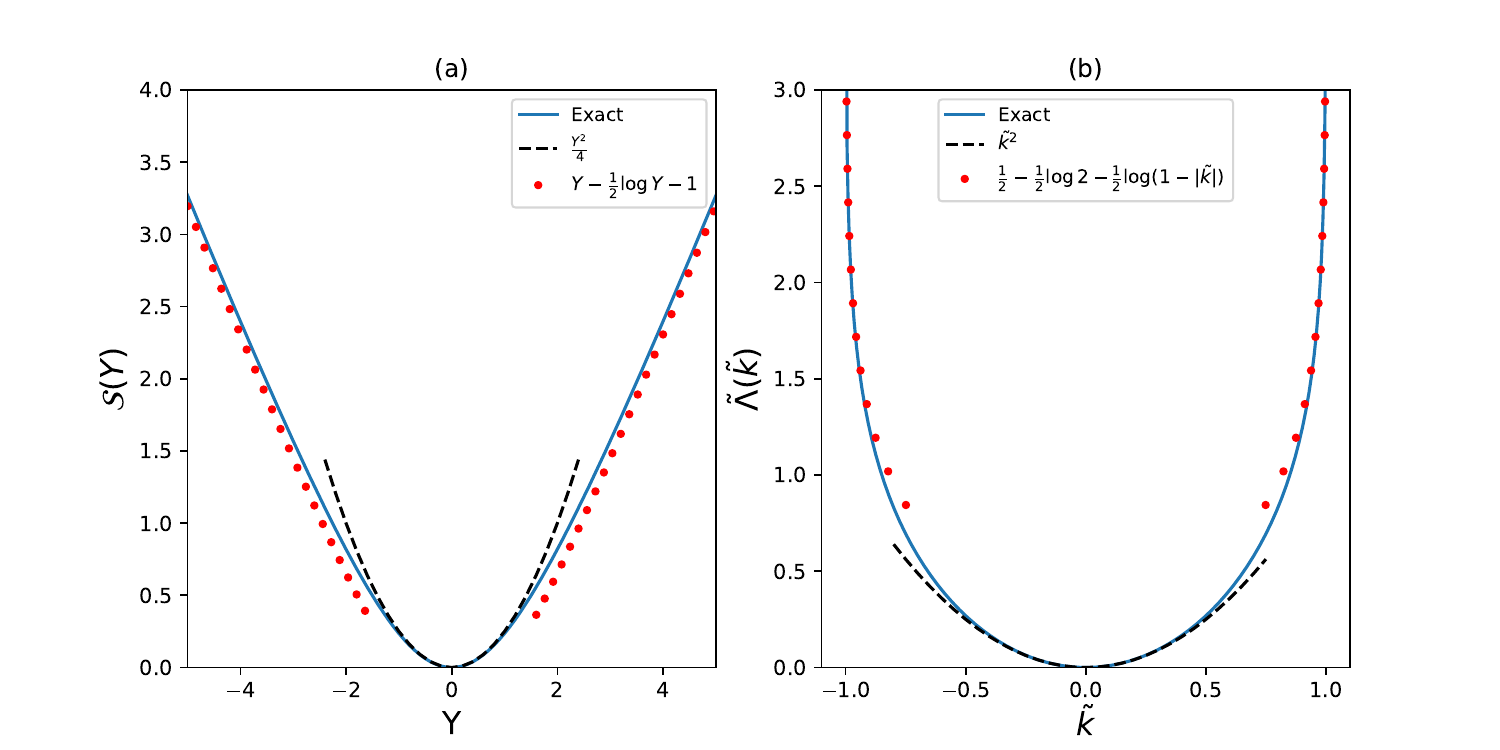}
    \caption{Plots of the scaled action $\mathcal{S}(Y)$ and SCGF $\tilde{\Lambda}(\tilde{k})$ for harmonic potential.  Here $D=\mu=1$.  The black dashed lines denote the Brownian approximations $\mathcal{S}(Y )\simeq \frac{Y^2}{4}$  for $Y \simeq 0$ and $\tilde{\Lambda}(\tilde{k}) \simeq \tilde{k}^2$  for $\tilde{k} \to 0$. The red dotted lines denote the asymptotic behaviors for  $|Y| \to \infty$ and $|\tilde{k}| \to 1$ (see Eq. \eqref{lambda asymp} and \eqref{action asymp}).}
    \label{rate func and scgf harmonic}
\end{figure}

  It is difficult to obtain an explicit expression for $\mathcal{S}(Y)$. However, as we now show, one can obtain an explicit expression for
$\tilde{\Lambda}(\tilde{k})$, the SCGF corresponding to $P_{\text{SSD}}(X)$: $\tilde{\Lambda}\left(\tilde{k}\right)=\lim_{\gamma\to\infty}\frac{\ln\left\langle e^{\sqrt{\gamma}\,\tilde{k} X}\right\rangle }{\gamma}$. 
$\tilde{\Lambda}(\tilde{k})$ is given by the Legendre(-Fenchel) transform of $\mathcal{S}(Y)$.
 Thus, inverting the relation $\tilde{k}=\tilde{k}(y) = \Omega(-y)$ (using \eqref{OmegaCubicEqn}) and using the Legendre transform relation, we obtain
\be \label{tilde k and y relation}
\frac{d\tilde{\Lambda}}{d\tilde{k}}=y=\frac{\tilde{k}^{3}-2\tilde{k}}{\tilde{k}^{2}-1} \, .
\ee
After integrating Eq. \eqref{tilde k and y relation} over $\tilde{k}$,  together with $\tilde{\Lambda}(0) = 0$, we obtain the SCGF
\begin{eqnarray} \label{scgf scaled}
    \tilde{\Lambda}(\tilde{k}) = \frac{\tilde{k}^2}{2} - \frac{1}{2}\,\,\ln (1-\tilde{k}^2) \, .
\end{eqnarray}
$\tilde{\Lambda}(\tilde{k}) $ is convex and differentiable function of $-1 \leq \tilde{k} \leq 1$ (see Fig. \ref{rate func and scgf harmonic}(\textbf{b})). $\mathcal{S}(Y)$ may then be obtained as the Legendre transform of $\tilde{\Lambda}(\tilde{k})$, giving us an alternative (parametric) representation to the one given above in \eqref{SYHarmonic1}:
\bea
Y&=&\frac{d\tilde{\Lambda}}{d\tilde{k}}=\tilde{k}+\frac{\tilde{k}}{1-\tilde{k}^{2}}=\frac{\left(2-\tilde{k}^{2}\right)\tilde{k}}{1-\tilde{k}^{2}} \, ,\\[1mm]
\mathcal{S}&=&\tilde{k}Y-\tilde{\Lambda}=\frac{1}{2}\left[\frac{\left(3-\tilde{k}^{2}\right)\tilde{k}^{2}}{1-\tilde{k}^{2}}+\log\left(1-\tilde{k}^{2}\right)\right] \, .
\eea
 In Appendix \ref{comparison harmonic}, we show that our result \eqref{scgf scaled} is in perfect agreement with the results of \cite{SDN2021}, in the limit $\gamma\to\infty$ (with constant $\tilde{k}$).

The asymptotic behaviours of $\tilde{\Lambda}(\tilde{k})$ are:
\be \label{lambda asymp}
\tilde{\Lambda}(\tilde{k}) \simeq \begin{cases}
\tilde{k}^2, & \tilde{k} \to 0,\\[2mm]
\frac{1}{2}-\frac{1}{2}\ln2-\frac{1}{2}\ln\left(1-|\tilde{k}|\right), & |\tilde{k}| \to 1 .
\end{cases}
\ee
The asymptotic behaviours of  $\mathcal{S}(Y)$ can be evaluated from the Legendre transformation of the asymptotic behaviours of $\tilde{\Lambda}(\tilde{k})$:
\be \label{action asymp}
\mathcal{S}(Y) = \begin{cases}
\frac{Y^2}{4} \, , & |Y| \ll 1 \, ,\\[2mm]
|Y| - \frac{1}{2}  \log{|Y|} -1 \, ,  & |Y| \gg 1 \, .
\end{cases}
\ee
Fig. \ref{rate func and scgf harmonic} shows the plot of the action $\mathcal{S}(Y)$ and the SCGF $\tilde{\Lambda}(\tilde{k})$, together with their asymptotic behaviors.


\begin{figure}
    \centering
\includegraphics[scale=0.85]{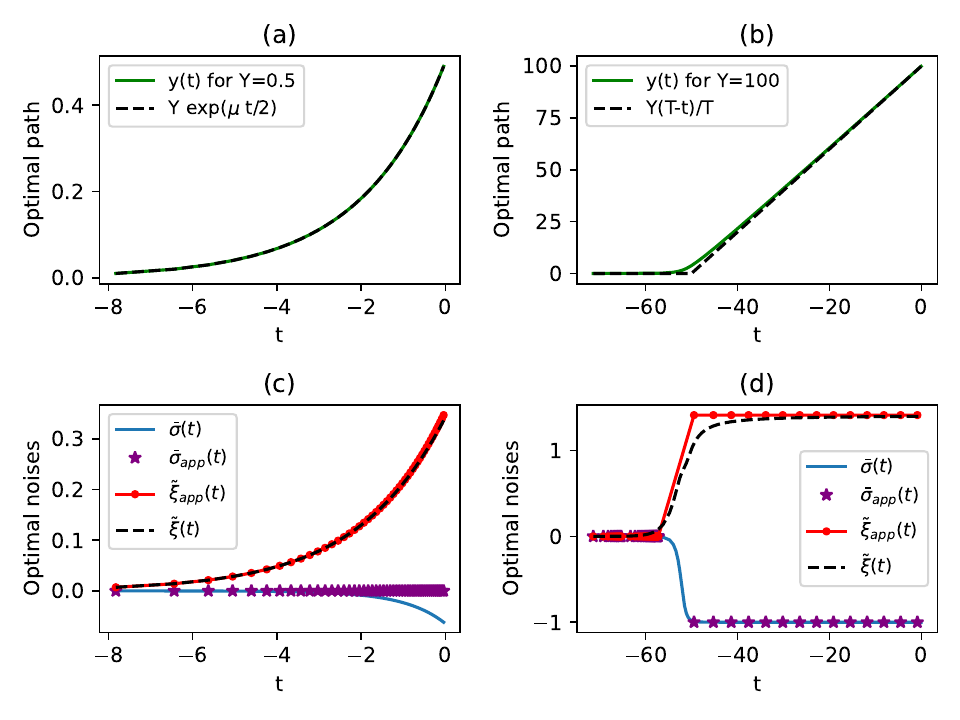}
\caption{Optimal trajectory $y(t)$, coarse grained telegraphic noise $\bar{\sigma}(t)$, and scaled noise  $\tilde{\xi}(t) =1/ \sqrt{\gamma} \, \xi(t) $ for small and large limits of $|Y|$ for harmonic potential. (a) and (c) denotes the plots for small-$Y$ regime (Fig. represents for $Y=0.5$). 
(b) and (d) represent the plots for $Y=100$.  Here $\gamma \to \infty$ and $D=\mu=1$. Solid lines denote the optimal trajectories and noises calculated using the Eqs. \eqref{op trajec}, \eqref{optimal xi}  and  \eqref{optimal noise}. The dashed line in (a) represents the time-reversed relaxation trajectory  $Y\, e^{ t/2}$ (Eq. \eqref{harmonic traj small Y}) and in (b) it denotes the optimal path calculated in the $|Y| \to \infty$ approximation given by  Eq. \eqref{harmonic traj large Y}. The dot-dashed and star lines in (c) and (d) depict the values of the noises $\bar{\sigma}(t)$ and $\tilde{\xi}(t)$ in the asymptotic limits $Y \to 0$ and $|Y| \to \infty$ given by Eq. \eqref{opt noise large X} and Eq. \eqref{opt noise small X} respectively.  
}
\label{optimal trajec and noise harmonic}
\end{figure}

\smallskip

 We now calculate the optimal (most likely)
coarse-grained trajectory and optimal realization of the noises of the system conditioned on a given value of $X$ (as shown in Sec. \ref{sec optimal path}) for harmonic potential. We use the rescaled optimal trajectory $y(t) = x(t) / \sqrt{\gamma}$, $Y = X / \sqrt{\gamma}$ and rescaled noise $\tilde{\xi}(t) = \xi(t) / \sqrt{\gamma}$ along with the Eqs. \eqref{op trajec}, \eqref{optimal xi} and \eqref{optimal noise} to calculate the optimal trajectory $y(t)$ and noises $\bar{\sigma}(t)$, $\tilde{\xi}(t)$  and then compare them with the asymptotic approximation \eqref{optimal path large X}, \eqref{activation trajec}, \eqref{opt noise large X} and \eqref{opt noise small X} (after rescaling). 
Fig. \ref{optimal trajec and noise harmonic} shows the plots for optimal trajectory and noises for $Y=0.5$ and $Y=100$.

Fig. \ref{optimal trajec and noise harmonic} (\textbf{a}) and (\textbf{c})  show the plots for the typical fluctuation regime (here $Y=0.5$). Similarly,  Fig. \ref{optimal trajec and noise harmonic} (\textbf{b}) and (\textbf{d})  show the plots for $|Y| \to \infty$ regime (here $Y=100$).  The optimal trajectory for $Y \simeq 0$ follows the Langevin dynamics given by Eq. \eqref{Effective Langevin} and it is the time reversed relaxation trajectory given by the solution of Eq.\eqref{activation trajec}:
\be \label{harmonic traj small Y}
y(t) = y_{\text{rel}}(-t) = Y \, e^{t/2}.
\ee
see Fig. \ref{optimal trajec and noise harmonic} (a). 
The optimal noises for this case are given by Eq. \eqref{opt noise large X} (see Fig. \ref{optimal trajec and noise harmonic}(c)). 
\be 
\bar{\sigma} \simeq 0 \quad \text{and} \quad
\tilde{\xi}(t) \simeq y(t) /\left(2\sqrt{2}\right).
\ee
Notably, the $|Y|\ll1$ approximations are very accurate even for the choice $Y=1/2$.

For $Y=100$, the optimal trajectory is given by Eq. \eqref{optimal path large X} (see Fig. \ref{optimal trajec and noise harmonic}(b))
\be  \label{harmonic traj large Y}
y\left(t\right)\simeq
\begin{cases}
0, & t<T,\\[2mm]
Y \, \Big(1 -  \frac{t}{T} \Big)\,, & T<t<0,
\end{cases}
\ee
with $T=- |Y|/2  $. The optimal noises are given by Eq. \eqref{opt noise large X} (see Fig. \ref{optimal trajec and noise harmonic}(d))
\be 
\begin{cases}
\bar{\sigma}(t) \simeq 0,\,\, \tilde{\xi}(t)= 0, \, & t<T,\\[1mm]
\bar{\sigma}(t) \simeq -1\,, \,\, \tilde{\xi}(t)= \sqrt{2 }, \, & T<t<0.
\end{cases}
\ee




\section{Dynamical phase transition for one-dimensional periodic systems} \label{sec generalization}

We now consider here a one-dimensional potential $U(x)$ with periodic boundary conditions.
 Let us assume that $U(x)$ has a global minimum at $x=x_0$ and global maximum at $x=x_1$, and for simplicity, assume that $U(x)$ is a smooth function and has no other local minima or maxima.
We choose units of distance such that the length of the system is $2\pi$, so that we will perform our analysis on the interval $x_1 < x < x_1 + 2\pi$ with periodic boundary conditions. Unlike the cases we treated above, in which the system was infinite, here $x$ is bounded and so is $U(x)$. 
 As a result, one cannot consider the large-$X$  limit (as we did, e.g., in the previous section), so the action $S$ that arises is in the OFM calculations is in general not large unless one additionally assumes that $D$ is small.
Therefore, we will use a different scaling in this section: We assume that $\gamma \to \infty, D \to 0$ but with their product $\tilde{D} = \gamma D$ constant.
Indeed, Eq.~\eqref{kF(x) relation} then becomes
$k = \frac{\gamma}{\sqrt{\tilde{D}}}\, \Omega\left(\frac{F\left(x\right)}{\sqrt{\tilde{D}}}\right)$
and then from Eq.~\eqref{action calculation} we find that  the action takes the scaling form
$S\left(X, \gamma, D\right) = \gamma \, s\left(X,\tilde{D}\right)$
where $s(X, \tilde{D})$ is the large-deviation function that describes the SSD through the LDP
$P(X;\gamma,D) \sim e^{-\gamma s(X,\tilde{D})}$.



\begin{figure}
\centering
     \begin{subfigure}[h]{0.41\textwidth}
         \centering
         \includegraphics[width=\textwidth]{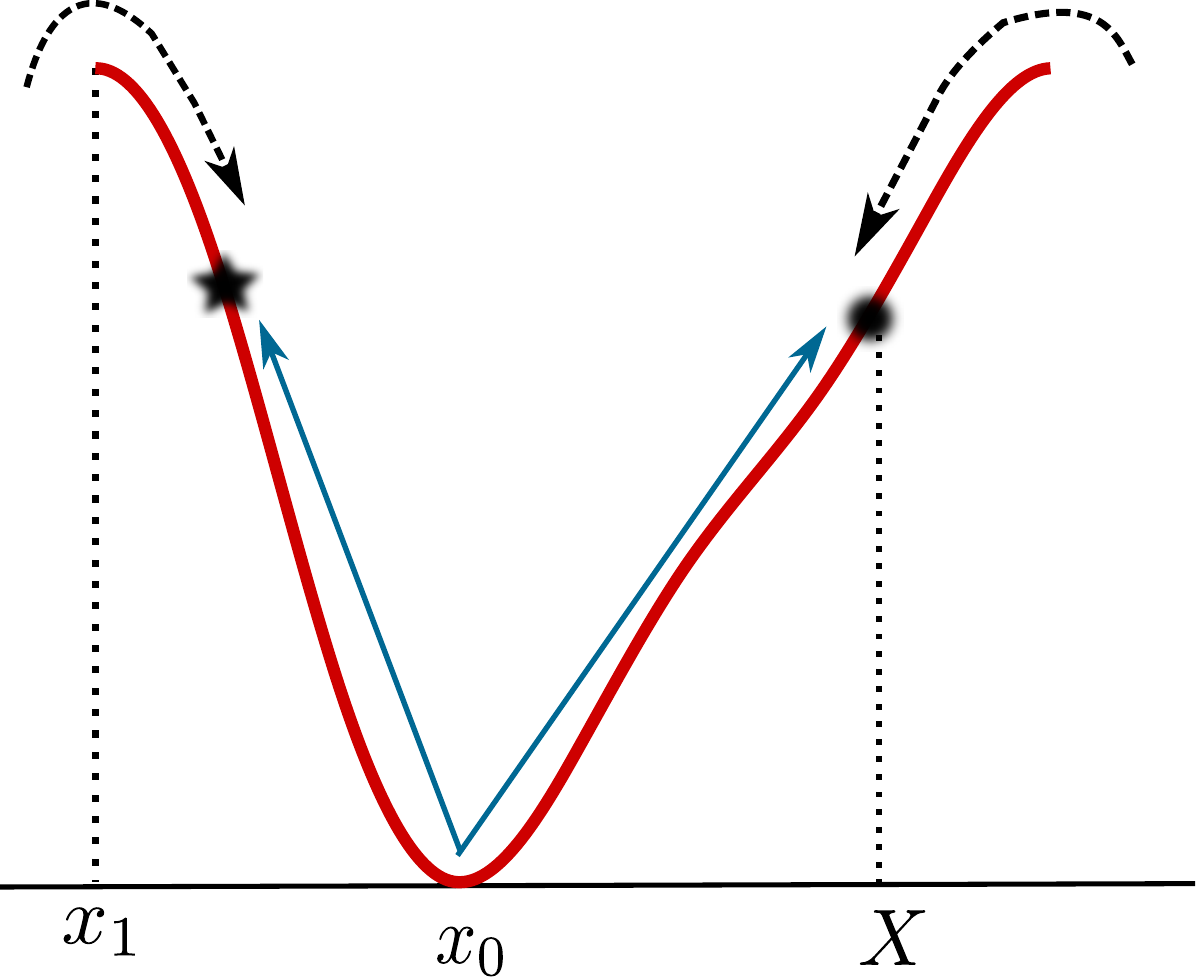}
         \caption{}
         \label{fig periodic pot}
     \end{subfigure}
     \hfill
     \centering
     \begin{subfigure}[h]{0.55\textwidth}
         \centering
         \includegraphics[width=\textwidth]{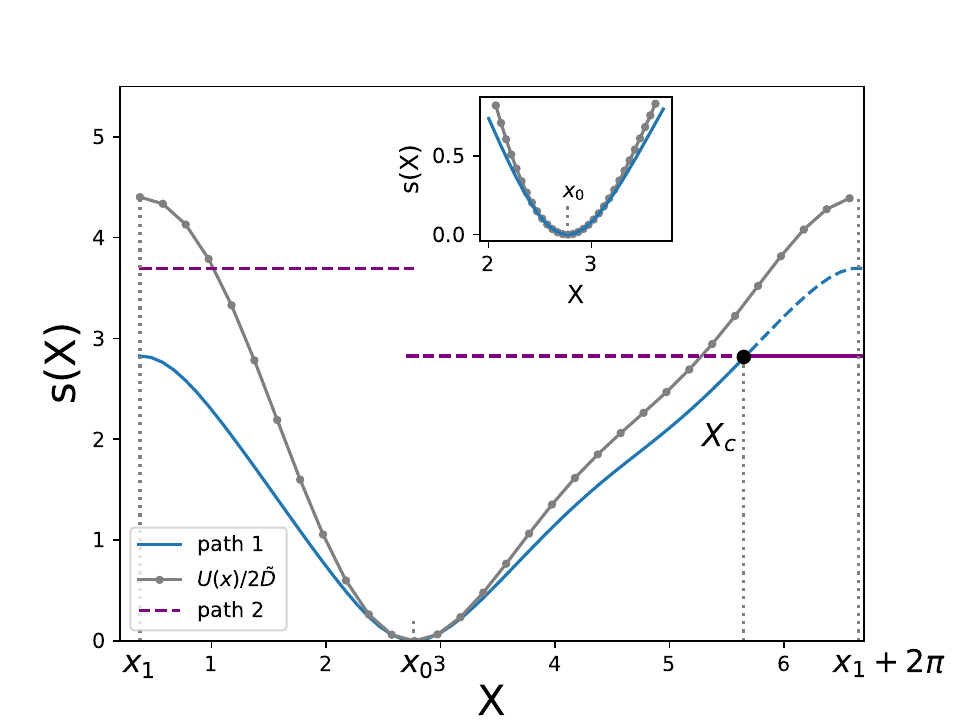}
         \caption{}
         \label{fig action periodic}
     \end{subfigure}
     \hfill
        \caption{ \textbf{(a)} Red solid line:  The trapping potential 
        \eqref{Uexample}. Here $x_0$, $x_1$ are the minimum and maximum of the potential,  respectively. Solid blue arrows denote the paths $1$ to reach the target positions, shown by solid circle and star. Dotted lines denote the paths $2$ for the same targets.  In the text we focus on the case $X>x_0$, corresponding to the circle. \textbf{(b)} Solid line: The scaled action $s(X)$ as a function of $X$ for periodic potential \eqref{Uexample}.   $s(X)$ is the minimum of the scaled actions of two paths leading from $x_0$ to $X$ (in the left and right directions). Dashed lines denote the  action of the non-optimal path. Dot-dashed line denotes $U(X)/2 \tilde{D}$, corresponding to  the effective Boltzmann  distribution that describes typical fluctuations. The inset shows the region where the effective Boltzmann distribution is valid. $s(X)$ exhibits a  singularity  at $X=X_c=5.365\dots$ (shown by black solid circle) at which the first derivative $ds(X)/dX$ shows a discontinuity. We interpret this as a first order dynamical phase transition.  Here $\tilde{D} = 0.25$. 
       }
        \label{fig periodic plot}
\end{figure}

 However, when evaluating the action $S(X)$ using Eq.~\eqref{action calculation},
due to periodicity, there are two paths to reach any $X \neq x_0$.
 For simplicity, let us analyze the case in which $x_0 < X < x_1 + 2\pi$, see Fig. \ref{fig periodic plot}.
The path $1$ is the trajectory starting from $x_0$ and moving right straight towards $X$,
 \be
S\left(X\right)_{\text{path} \,1}=\int_{x_{0}}^{X}\frac{\gamma}{\sqrt{\tilde{D}}}\, \Omega\left(\frac{F\left(x\right)}{\sqrt{\tilde{D}}}\right)\, dx \, .
 \ee
 The other path (path $2$) is the trajectory starting from $x_0$ moving towards left to $x_1$ and then from $x_1 + 2 \pi$ to $X$.  The action corresponding to this path is
 \be
S\left(X\right)_{\text{path 2}}=\int_{x_{0}}^{x_1}\frac{\gamma}{\sqrt{\tilde{D}}}\, \Omega\left(\frac{F\left(x\right)}{\sqrt{\tilde{D}}}\right)\, dx \, .
 \ee
The contribution of the part of the trajectory from $x_1 + 2 \pi$ to $X$ in the left direction vanishes; The particle simply rolls ``downhill'' due to the external potential. 
The  (minus) logarithm of the SSD is hence given by the minimum of these two actions:
\be
S(X)=\min\left\{ S\left(X\right)_{\text{path}\,1},S\left(X\right)_{\text{path}\,2}\right\} \, .
\ee
The contribution of the non-optimal path to $P_{\text{SSD}}(X)$ is negligible in the limit of small $D$ and large $\gamma$.

 For a generic periodic potential $U(x)$  (under the assumptions stated above), this leads to the emergence of a remarkable behavior. Namely, as there may exist a critical value of $X=X_c$ at which a switching happens between  the optimality of $S\left(X\right)_{\text{path} \,1}$ and $S\left(X\right)_{\text{path} \,2}$. This switching behaviour leads to a non-differentiable point in $S(X)$ at $X=X_c$, which we interpret as dynamical phase transition (DPT). The first derivative of $S(X)$ exhibits a discontinuity at $X_c$, hence it is a first order DPT.
 It is not difficult to see that if the action along the route from $x_0$ to $x_1$ in the left direction is smaller than the one from $x_0$ to $x_1+2\pi$ (in the right direction), i.e., if $S_2 < S_1$ where
$S_2 = \int_{x_{0}}^{x_1}k dx$
and $S_1 = \int_{x_{0}}^{x_1 + 2\pi}kdx$,
then the DPT occurs at a critical value $X_c \in (x_0,x_1+2\pi)$. In the converse case $S_2 > S_1$, the DPT occurs at a value $X_c \in (x_1, x_0)$.
Finally, if the potential is mirror symmetric  around $x_0$, $U(x) = U(x_0-x)$, then $S_1 = S_2$ and no DPT occurs.
 Naturally, since the scaled action $s$ is simply given by $s=S/\gamma$, the above claims regarding a DPT hold for $s$ as well.

We  illustrate these results by considering 
\be
\label{Uexample}
U(x) = \cos x + (1/4) \sin 2x  \, + U_0
\ee
to be an example (see Fig. \ref{fig periodic pot}). Here $x_0 = \pi -\tan ^{-1}\left(\sqrt{\frac{2}{\sqrt{3}}-1}\right) \simeq 2.766\dots$ is the minimum of the potential, $x_1= \tan ^{-1}\left(\sqrt{\frac{2}{\sqrt{3}}-1}\right) \simeq 0.374\dots$ is the maximum of the potential,  and $U_0$ is a constant that is added in order to ensure that $U(x_0)$ = 0 (for convenience).
Fig. \ref{fig action periodic} shows the plot of $s(X)$ (shown in solid line) for $\tilde{D} = 0.25$.
 Typical fluctuations are described by the effective Boltzmann distribution, which corresponds to the behavior  $s(X) \simeq U(X)/(2 \tilde{D})$ for $X \simeq x_0$.
 For our example \eqref{Uexample}, $S_1 > S_2$,  and, as a result, a DPT occurs at a critical value $X_c \in [x_0, x_1 + 2\pi]$ which we find shortly.
For $X > x_0$ (shown by  the black circle in Fig.~\ref{fig periodic pot}), the path $1$ (shown by  the solid blue line   with the arrow pointing to the right at $X$  in Fig.~\ref{fig periodic pot})  is optimal until $X < X_c$. Above $X > X_c$, path $2$ (shown by the dotted line  with the arrow pointing towards the black circle in Fig.~\ref{fig periodic pot}) is optimal. $s(X)$ hence exhibits a first order DPT at $X=X_c$, shown by black dot in Fig. \ref{fig action periodic}. $X_c$ can be calculated using the relation:
\be
S(X_c)_{\text{path 1}} = S(X_c)_{\text{path 2}} \implies X_c = 5.365\dots.
\ee
 At $x_1 < X < x_0$ (shown by black star in Fig.~\ref{fig periodic pot}), the direct trajectory from $x_0$ to $X$ (shown by solid blue line towards left in Fig.~\ref{fig periodic pot}) in the left direction is always optimal (since $S_1 > S_2$ in this example).

 A qualitatively similar  DPT occurs   for Brownian particle in a standard (non-intermittent) periodic potential,  but only in the presence of an external drift  \cite{Baek15,GT86}. In that case the first order DPT is related to the breaking of time-reversal symmetry due to the external drift. In our system  we did not assume an external drift, and the DPT reflects the intrinsic nonequilibrium behavior of the dynamics  due to the intermittency of the external potential.  
 
 An additional nonequilibrium effect we observe is a nonzero steady-state probability current $j \ne 0$ (which, for Brownian particles in the absence of intermittency, is forbidden). 
  In the leading order, the probability current is given by $j \sim p_1 - p_2$ where $p_{1,2} \sim e^{-S_{1,2}}$ are the probabilities for the particle to climb from the minimum of $U(x)$ to the maximum of $U(x)$ in the left and right directions, respectively, and $j>0$ ($j<0$) describes current in the right (left) direction. For $\gamma \to \infty$ and $D \to 0$, these two probabilities in general differ by many orders of magnitude, leading to
 \be
 j\sim\begin{cases}
e^{-S_{1}} \, , & S_{1}<S_{2} \, ,\\[1mm]
-e^{-S_{2}} \, , & S_{1}>S_{2} \,.
\end{cases}
\ee
 Since the actions $S_i$ are proportional to the switching rate $\gamma$, and we are considering the large $\gamma$ limit, the current $j$ is very small, and vanishes%
\footnote{ Note that, in the scaling limit considered in this section, the vanishing of $j$ may be jointly attributed to two different effects:
(i) At large $\gamma$, the dynamics may be approximated by the effective dynamics \eqref{Effective Langevin} which correspond to a system that is in equilibrium.
(ii) At small $D$, hopping events between consecutive minima become very unlikely, and at $D \to 0$ they cannot occur.
We do not attempt here to study each of these to effects separately, but rather we consider the joint limit $\gamma \to \infty, D \to 0$ with constant $\tilde{D} = \gamma D$.}
 at $\gamma \to \infty$.
The large difference between the orders of magnitude of $p_1$ and $p_2$ can be exploited in order to separate between diffusing particles of two different types, if $S_1 < S_2$ for one type of particle and $S_1 > S_2$ for the other type, cf. \cite{HAAS23}.
If $U(x)$ is mirror symmetric around $x_0$, then $j=0$ (exactly) due to the symmetry.
 A nonzero steady-state probability current in the absence of an external drift can also occur for active particles trapped in periodic potentials (without intermittency) \cite{led20, ACD11, MP21, MartinEtAl21, OKTW22}. There it is the activity which breaks time-reversal symmetry.
  More generally, these are all examples of the ``ratchet principle'', which states that in non-equilibrium systems and in the absence of mirror symmetry, there will in general be nonzero steady-state currents \cite{Magnasco93, Magnasco94, AMPP94}.

\bigskip


 \section{Conclusion} \label{sec conclusion}

In this paper we studied a Brownian particle  in $d$ dimensions under the effect of an intermittent potential, which switches on and off at a constant rate $\gamma$. We considered a smooth potential with a single minimum,  and for $d>1$ we additionally assumed the potential to be central. 
The intermittency of the potential breaks the time-reversal symmetry of the system and leads to nonequilibrium behavior. 
We studied the steady state position distribution $P_{\text{SSD}}(\boldsymbol{X})$ that is reached at long times, and the MFPT to reach a given point,  focusing on the rapid-switching limit $\gamma \to \infty$.

For $\boldsymbol{X}$ that is sufficiently close to the center of the potential,  $P_{\text{SSD}}(\boldsymbol{X})$ follows an effective Boltzmann distribution and the MFPT follows a corresponding Arrhenius law. However, 
at large $X = |\boldsymbol{X}|$, the distribution $P_{\text{SSD}}(\boldsymbol{X})$ deviates from this Boltzmann behavior, and the MFPTs deviate from the Arrhenius law, cf. \cite{NF2022, NS2023, GMS24}.
We obtained the full distribution $P_{\text{SSD}}(X)$  in $d=1$ by developing an effective coarse-grained description of the system's dynamics, using the temporal additivity principle 
\cite{NF2022, NS2023, HT2009, Harris2015, Jack2019, JH2020, AB2021, NF2022, NS2023, BTZ2023, MS2024, PLC24, DKB25}.
Besides the SSD itself, our formalism also yields the most likely (coarse-grained) history of the system conditioned on observing a given value of $X$.
In dimensions $d>1$ with a central potential $U(\boldsymbol{x}) = U(x)$, we find that in the leading order, the SSD behaves as
$P_{\text{SSD}}(\boldsymbol{X}) = P^{1 \text{D}}_{\text{SSD}}(X)$
where
$P^{1 \text{D}}_{\text{SSD}}(X)$
is the SSD for the case $d=1$ with confining potential
$U(x)$.
Remarkably, we found that the behavior of $P_{\text{SSD}}(\boldsymbol{X})$ in the far tails, $|\boldsymbol{X}| \to \infty$, is universal 
for any smooth confining potential, and is described by decaying exponentials.  Interestingly, these tails coincide, in the leading order, with those of the SSD of a Brownian motion with stochastic, instantaneous resetting to the origin at  rate $\gamma$. We uncovered the physical mechanism behind this coincidence, by showing that the optimal histories in the two systems, conditioned on reaching a given point, coincide as well. 
Furthermore, universal exponential tails have also been observed in many other systems which do not involve intermittent potentials \cite{BB20, WBB20, HuEtAl23, Burov20, SB23, HBB24,  SB24, CK20, SIJ23}.

We next studied periodic one dimensional systems with an intermittent potential. In the limit of $D \to 0$ and $\gamma \to \infty$, we found that the logarithm of $P_{\text{SSD}}(X)$ develops a singularity at a critical point $X=X_c$, which we interpreted as a DPT of first order. Our formalism reveals that the DPT  follows from the fact that there are two paths leading from the minimum of the potential to any point $X$.
For a Brownian particle in a standard (non-intermittent) potential, a similar DPT occurs, but only if an additional external drift is applied \cite{Baek15,GT86}. For intermittent potentials, the DPT occurs in general even without applying such a drift.
In addition, we calculated the nonequilibrium probability current of the system in its steady state.


While we have derived exact results (in the limit $\gamma \to \infty$) for the case where the potential is smooth and has a single global minimum, the framework developed here can readily be extended to more complex confining
potentials (e.g., double-well potentials). There are several open questions and future directions related to our work. Our theoretical framework can be extended to more general dynamics for the potential's intensity $\eta$ (e.g., not restricted to only $\eta = 0, 1$ corresponding to off and on states, respectively, and/or non-Markovian $\eta(t)$) \cite{BKMS24, Frydel24}. It would be interesting to calculate the steady state distribution in higher dimensions for general (non-central) confining intermittent potentials. Since there are multiple paths leading to each point $\boldsymbol{X}$, we expect DPTs to occur, like in the periodic 1D system that we studied here. Another interesting direction, which we expect could be analyzed using a similar approach to the one we used here, could be to study Brownian motion with a fluctuating diffusion coefficient, trapped in an external potential \cite{MUA2019, GMS25}.

\bigskip

\section*{Acknowledgments}

NRS acknowledges support from the Israel Science Foundation (ISF) through Grant No. 2651/23, and from the Golda Meir fellowship.

\bigskip


\begin{appendices}

\section{Calculation of the scaled cumulant generating function} \label{SCGFcalculation}

\renewcommand{\theequation}{A\arabic{equation}}
\setcounter{equation}{0}

Here, for completeness we calculate explicitly the SCGF corresponding to the long-time distribution of the position $\tilde{x}(t)$, as defined in Eq. \eqref{coarsegrainedSDE const f}, which, for convenience we give here again:
\begin{eqnarray} \label{tilde x}
    \tilde{x}(t)=\frac{f \, t}{2}+\frac{f}{2} \int_0^t \, \sigma \left(t'\right) \, dt' + \sqrt{2D}\, \int_0^t \, \xi\left(t'\right) \, dt'.
\end{eqnarray}
This corresponds to a constant force $F(x)=f$.
At long times, the position follows an LDP:
\be 
P[\tilde{x}(t)] \sim e^{- t\,  \Psi_f(\tilde{x}/t)} \, .
\ee
The corresponding SCGF for $\tilde{x}$ is 
\be
\lambda_{f}\left(k\right)=\lim_{t\to\infty} \frac{1}{t} \, \ln\langle e^{k\, \tilde{x}}\rangle \, .
\ee
Plugging Eq. \eqref{tilde x} in $\lambda_{f}\left(k\right)$, we get 
\be
\lambda_{f}\left(k\right) = \lim_{t\to\infty} \frac{1}{t} \, \ln \Bigg \langle e^{k\, \Big (\frac{f \, t}{2}+\frac{f}{2} \int_0^t \, \sigma \left(t'\right) \, dt' + \sqrt{2D}\, \int_0^t \, \xi\left(t'\right) \, dt' \Big)} \Bigg \rangle \, .
\ee
Since the cumulants of the sum of i.i.d random variables are the sum of their cumulants, $\lambda_{f}\left(k\right)$ becomes the sum of the SCGFs of each of the terms in the right hand side of Eq.~\eqref{tilde x},
\be
\label{lambdafxtilde}
\lambda_{f}\left(k\right) = \lim_{t\to\infty}  \frac{1}{t}  \Bigg [ \frac{k \, f}{2} \, t +  \ln \Big \langle e^{\frac{f \, k}{2} \int_0^t \, \sigma \left(t'\right) \, dt'} \Big \rangle + \ln \Big \langle e^{\sqrt{2D}\, k \,  \int_0^t \, \xi\left(t'\right) \, dt'} \Big \rangle \Bigg ] \, .
\ee
The first term contributes to a constant term $f \, k / 2$. The third term comes from the  white noise $\xi(t)$ which follows the distribution
\be \label{pdf of xi}
P\left[\xi\left(t\right)\right]\sim e^{-\frac{1}{2}\,\int_{0}^{t}\xi^{2} \, dt'},
\ee
giving the contribution 
\be
\lim_{t\to\infty}  \frac{1}{t} \, \ln \Big \langle e^{\sqrt{2D}\, k \,  \int_0^t \, \xi\left(t'\right) \, dt'} \Big \rangle = D \, k^2 \, .
\ee
The SCGF corresponding to the second term in \eqref{lambdafxtilde} has been well known \cite{NF2022, NS2023}:  $ \sigma = \int_0^t \, \sigma \left(t'\right) \, dt'$  satisfies the LDP $\psi\left(\sigma \right)$:
\begin{eqnarray} 
P\left[ \sigma \right]\sim e^{- t \, \psi\left(\sigma \right)},
\end{eqnarray}
with rate function $\psi(\sigma) = \gamma - \sqrt{\sigma^2 + \gamma^2} $. 
The corresponding SCGF (which may be obtained from the Legendre transform of $\psi$) is
\be
\lim_{t\to\infty}  \frac{1}{t} \, \ln \Big \langle e^{\frac{f \, k}{2} \int_0^t \, \sigma \left(t'\right) \, dt'} \Big \rangle = \sqrt{\frac{f^{2}k^{2}}{4}+\gamma^2} -\gamma \, .
\ee
Hence, we get the SCGF that describes the long time distribution of $\tilde{x}$ is
 \begin{eqnarray} 
     \lambda_f(k) = \frac{f \, k}{2} + \sqrt{\frac{f^{2}k^{2}}{4}+\gamma^2} -\gamma +D\, k^{2}
 \end{eqnarray}
 which is Eq. \eqref{SCGF} in Sec. \ref{sec coarse grain}.

\bigskip

 \section{Comparison with exact results for the harmonic potential} \label{comparison harmonic}

\renewcommand{\theequation}{B\arabic{equation}}
\setcounter{equation}{0}

For the harmonic trapping potential, $U(x) = \mu x^2 /2$, we can compare our results with those of \cite{SDN2021}, providing a useful check. Eq. (22) of \cite{SDN2021} gives the exact Fourier transform 
\be
\hat{P}\left(k\right)=\int_{-\infty}^{\infty}P_{\text{SSD}}\left(x\right)e^{-ikx}dx
\ee
of the steady-state distribution $P_{\text{SSD}}(x)$:
\be
\hat{P}\left(k\right)=\frac{e^{-Dk^{2}/2\mu}}{2\left(1+Dk^{2}/\gamma\right)^{\gamma/2\mu}}\left(1+\frac{1}{1+Dk^{2}/\gamma}\right) \, .
\ee
By analytic continuation to the complex plane this implies that the moment generating function, i.e., the two-sided Laplace transform
\be
\left\langle e^{\kappa x}\right\rangle =\int_{-\infty}^{\infty}P_{\text{SSD}}\left(x\right)e^{\kappa x}dx
\ee
is given, in our units $D=\mu=1$, by
\be
\left\langle e^{\kappa x}\right\rangle =\hat{P}\left(k=i\kappa\right)=\frac{e^{\kappa^{2}/2}}{2\left(1-\kappa^{2}/\gamma\right)^{\gamma/2}}\left(1+\frac{1}{1-\kappa^{2}/\gamma}\right)\,.
\ee
From here, one can extract the SCGF directly:
\be
\lim_{\gamma\to\infty}\frac{\ln\left\langle e^{\sqrt{\gamma}\,\tilde{k}x}\right\rangle }{\gamma}=\lim_{\gamma\to\infty}\left[\frac{\tilde{k}^{2}}{2}-\frac{1}{2}\ln\left(1-\tilde{k}^{2}\right)+\frac{1}{\gamma}\ln\left(\frac{1+\frac{1}{1-\tilde{k}^{2}}}{2}\right)\right]=\frac{\tilde{k}^{2}}{2}-\frac{1}{2}\ln\left(1-\tilde{k}^{2}\right)\,,
\ee
and the result is indeed in perfect agreement with our Eq.~\eqref{scgf scaled}.

\end{appendices}

\end{document}